\documentclass[doublecol]{epl2}

\usepackage{amsmath, amssymb, graphics, graphicx}
\usepackage{textcomp, subfigure}\usepackage{color}

\usepackage{sistyle}					

\newcommand{\SHO}{\mathrm{SHO}} 
\newcommand{\CHO}{\mathrm{CHO}}
\newcommand{\meas}{\mathrm{meas}}

\definecolor{linkcolor}{rgb}{0,0,0.6} 
\usepackage[pdftex,colorlinks=true, pdfstartview=FitV, linkcolor= linkcolor, citecolor= linkcolor, urlcolor= linkcolor, hyperindex=true,hyperfigures=true]{hyperref} 

\begin{document}
\title{\bf Microrheology measurements with a hanging-fiber AFM probe}
\author{C. Devailly, J. Laurent, A. Steinberger, L. Bellon, S. Ciliberto }
\institute{
Universit\'e de Lyon \\
{Laboratoire de Physique, \'Ecole Normale Sup\'erieure de Lyon, CNRS}\\
46, All\'ee d'Italie, 69364 Lyon Cedex 07, France\\}
\date{today}
\begin{abstract}
{A method to measure the viscosity of liquids at microscales is presented. It uses a thin glass fiber fixed on the tip of the cantilever of an extremely low noise Atomic Force Microscope (AFM), which accurately measures the cantilever {deflection}. When the fiber is dipped into the liquid the dissipation of the cantilever-fiber system increases. This dissipation, linked to the liquid viscosity, is computed from the power spectral density of the thermal fluctuations of the cantilever {deflection}. The {high sensitivity of the AFM} allows us to show the existence and to develop a model of the coupling between the dynamics of the fiber and that of the cantilever. This model {accurately} fits the experimental data. The advantages and draw-backs of the method are discussed.
}
\end{abstract}
\maketitle


\section{Introduction}
The development of the study of complex fluids needs measurements at micrometer scale to observe local properties of {the} media. Furthermore one often needs to measure the viscosity of very small liquid samples. For these reasons several microrheology techniques have been developed. The most common is based on the measurement of the Brownian motion properties either by tracking several free micrometer beads or by trapping them in optical tweezers~\cite{Gittes,Mason,Solano}. Both techniques need a transparent fluid but an alternative method has been recently proposed in ref.~\cite{SMR_viscosity}. This method is based on an excited suspended microchannel resonator and it works for {viscosities} less than \SI{10}{mPa.s}. In ref.~\cite{Tong} {another technique has been proposed}, which measures the thermal fluctuations of a hanging-fiber Atomic Force Microscope (AFM) probe. This method is interesting and presents several advantages with respect to the other ones: a) It can work with large viscosities; b) the fluid has not to be transparent; c) the amount of needed fluid can be very small. {However this technique has been analyzed only around the system's resonance because of the limited sensitivity of the apparatus}. 

In this paper we analyze the advantages and drawbacks of this method with a very low-noise AFM. This improved sensitivity allows us to measure relatively high viscosities and most importantly to develop a model for the fiber-fluid interaction, which is more appropriate and precise than that of ref.~\cite{Tong}. 
Thus the results of this paper are not only useful for {microrheology} measurements of rather viscous and opaque fluids but also to {investigate} the basic dissipation mechanisms of a thin fiber inside a fluid and to give more insights to the mechanical coupling of {such} microdevices.

\begin{figure}[htb]
\begin{center}
\includegraphics[width=0.6\linewidth]{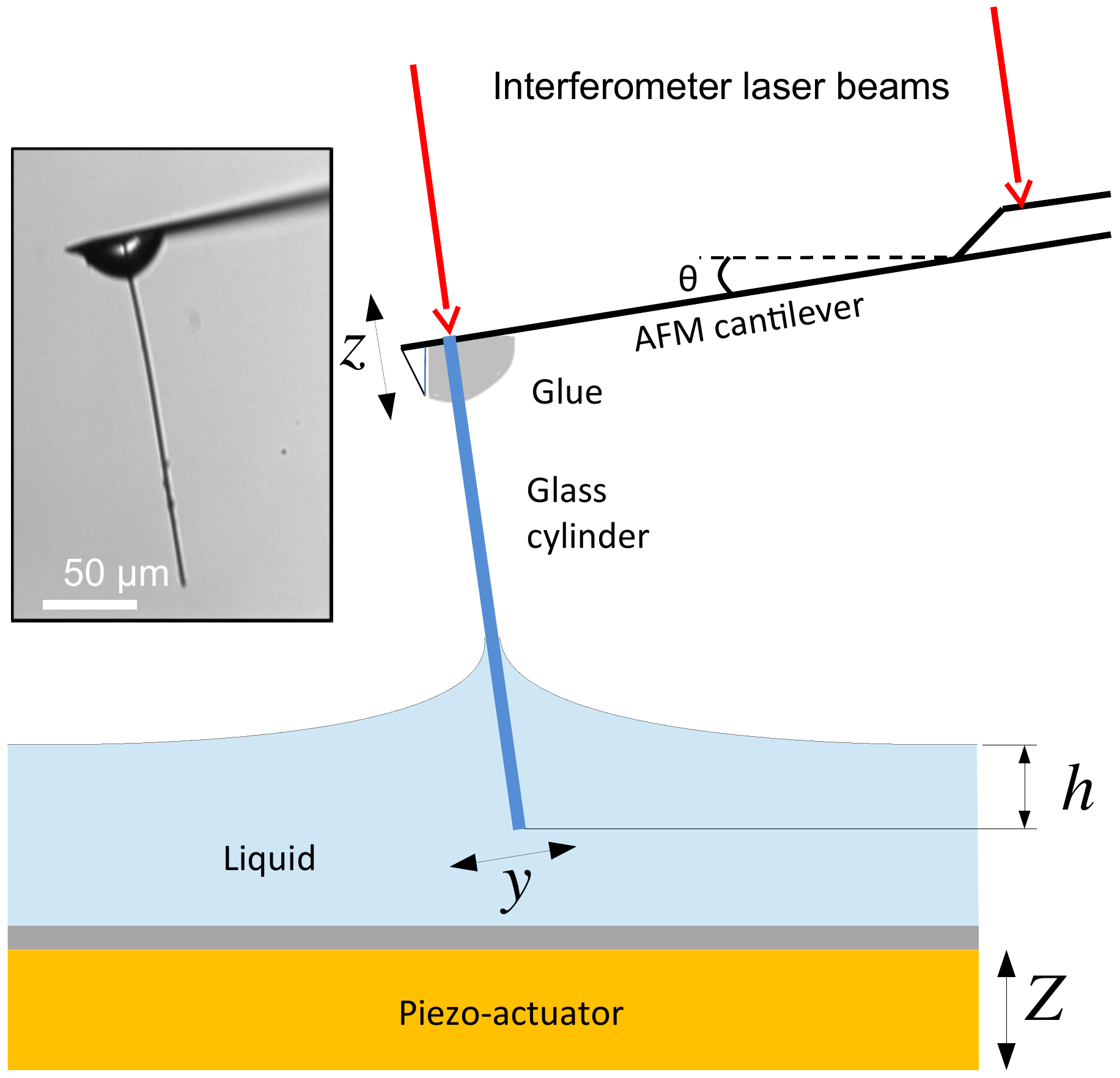}
\end{center}
\caption{Experimental set-up. The fiber {glued to} the cantilever is dipped in a liquid layer {on a depth $h$, controlled by the position $Z$ of the piezo-actuator }. The two laser beams of the interferometer measure the cantilever deflection {$z$. The fiber deflection is noted $y$. The} picture on the left is an image of the fiber{taken} with a bright field microscope x10 magnification.}
\label{fig:setup}
\end{figure}

\section{Experimental set-up}
The {measurement} of the fluid viscosity is performed using a cylindrical {micro-rod fixed to an AFM cantilever, as sketched in fig.\ref{fig:setup}. The rod} is immersed into the fluid whose friction along {its length} produces an extra damping for the cantilever, which
{is detected by measuring the thermal noise of the cantilever} 

{The rod is fabricated using an optical single-mode fiber, stretched under {the flame of a blowtorch}: its initial $\SI{125}{\micro m}$ diameter is thinned to $d\approx\SI{3}{\micro m}$. The fiber} is glued at the apex of {an} AFM cantilever with {a two-components epoxy adhesive (Araldite)}. {We use standard AFM cantilevers (Budget Sensors AIO): soft ones with a stiffness $k_{c}\approx\SI{0.25}{N/m}$ and a resonance frequency $f_{0}\approx\SI{8}{kHz}$, and stiffer ones with $k_{c}\approx\SI{4}{N/m}$ and $f_{0}\approx\SI{60}{kHz}$ (resonant frequencies after functionalization of the tip).} A particular effort is done to glue the fiber {as perpendicular to the cantilever as possible} to avoid torsion modes. The fiber is cut at the appropriate length {(around  $\SI{200}{\micro m}$) using a {sharp} tweezer} and a diamond tool. These operations are performed under a bright light microscope {with a x20 magnification,} using micromanipulators {(from Narishige)}. {Finally the fiber characteristics are checked with the {microscope} images (fig.\ref{fig:setup})}.

\begin{figure}[htb]
\begin{center}
\includegraphics[width=0.8\linewidth]{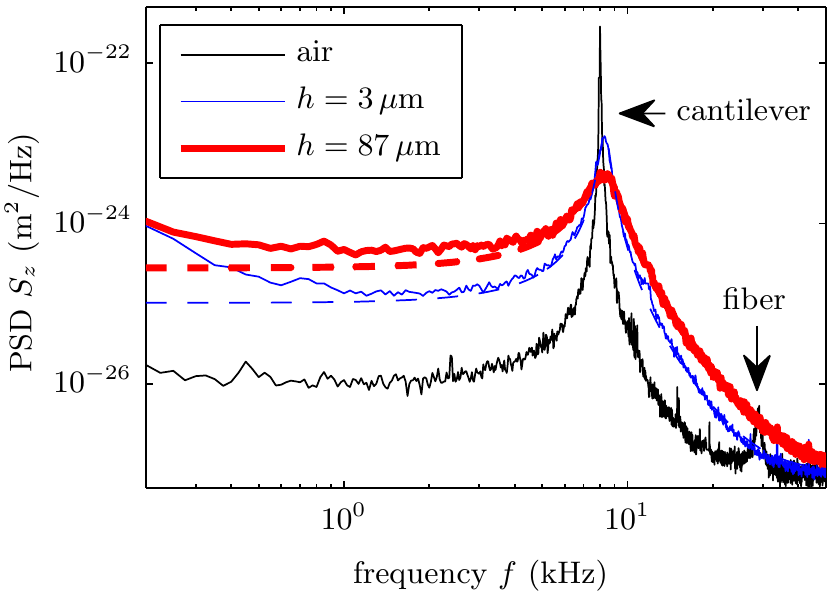}
\caption{{Soft cantilever, hexadecane. Solid line: measured PSD $S_{z}^{\meas}$ for several dipping depths $h$. Dashed line: fit with the spectrum $S_{z,0}^{\SHO}$ of the SHO model using only the data around the resonance (see text for details).}}\label{profondeurs}.
\end{center}
\end{figure}

{The hanging-fiber probe is mounted on} a home-built atomic force microscope. 
The deflection $z$ of the cantilever (figure \ref{fig:setup}) is measured by an interferometric deflection sensor~\cite{Paolino-2013}, inspired by the original design of Schonenberger~\cite{Schonenberg-RSI-89} with a quadrature phase detection technique~\cite{2002Bellon}: the interference between the reference laser beam reflecting on the {static base} and the sensing beam on the {tip of the} cantilever gives a direct measurement of the deflection with {a} very high accuracy (see spectra fig.\ref{profondeurs}). This technique offers a very low intrinsic noise (down to $\SI{E-14}{m/\sqrt{Hz}}$ {~\cite{Paolino-2013}}) and it is intrinsically calibrated. Indeed, contrary to the standard optical lever technique, measuring the angular deflection of the laser beam, the interferometric techniques measures directly the cantilever {deflection in term of the laser wavelength. Thanks to} this high resolution, no external excitation of the cantilever is {required, and its thermal fluctuations can be measured over a wide frequency range}.

The {viscosity} {measurement} is performed using {the same fiber with} {4} different liquids: alcanes {(dodecane and hexadecane) and silicon oils (10v and 20v, respectively of viscosity \SI{10}{mPa.s} and \SI{20}{mPa.s})} chosen for their range of viscosity, their weak evaporation, and their {good wetting of the glass fiber. The liquid is put in a $\SI{1.4}{cm}$ diameter copper container placed under the fiber; the liquid layer is about $\SI{1}{mm}$ deep. When changing the liquid, the fiber is rinsed thoroughly with the new liquid, and the copper container is rinsed as well before putting fresh liquid in it.  Silicon oils have been studied after alcanes, in an increasing order of viscosity.}
The pool can be moved along the {vertical axis $Z$} by a piezo-actuator. 
{The measurements are performed at $25^0C$}

In order to change the  {vertical height $h$ between the tip of the fiber and the undisturbed surface of the liquid (see fig.~\ref{fig:setup})}, the {position $Z$} of the {container} is changed in $\SI{3}{\micro m}$ steps. We wait { a few seconds} after {the} displacement before {starting the} measurement to let the fluid relax.
 {The origin of the height $h$ is located with a $\SI{3}{\micro m}$ uncertainty either from the jump of the static deflection due to the capillary force on the fiber or from the sudden broadening of the noise spectra between two displacement steps. The error on the relative height results from the mechanical drift of the actuation mechanism and the evaporation of the liquid, and is estimated to be lower than $\SI{5}{\micro m}$  for a typical $\SI{100}{\micro m}$ dipping experiment.}

{The cantilever deflection is sampled for at least  $\SI{5}{s}$, with a 24bits resolution,  at $\SI{240}{kHz}$ for the soft cantilever and $\SI{500}{kHz}$ for the stiff one.} The Power Spectral Density (PSD) of the deflection is calculated with a resolution of about $\SI{100}{Hz}$ for {the data in liquid and $\SI{25}{Hz}$ for data in air}. The spectra are averaged more than {$1000$} times in order to reduce the statistical noise. 
\begin{figure}[htb]
\begin{center}
\includegraphics[width=0.8\linewidth]{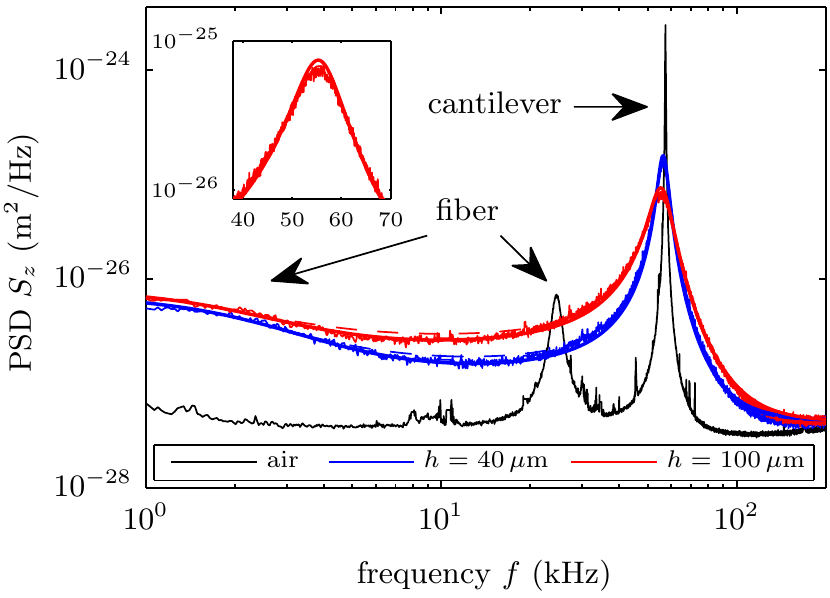}
\caption{{Stiff cantilever, hexadecane. Solid line: measured PSD $S_{z}^{\meas}$ for several dipping depths $h$. Dashed line: fit with $S_{z,0}^{\CHO}$. Bold solid line: fit with $S_{z,B}^{\CHO}$. Black: air, blue: $h=\SI{40}{\mu m}$, red: $h=\SI{100}{\mu m}$.}\label{HFhexadecane}}
\end{center}
\end{figure}
\section{Data analysis and parameters definition}
{In figs.~\ref{profondeurs} and \ref{HFhexadecane} the {measured PSD $S_{z}^\meas$ }of the cantilever deflection induced by the thermal noise is plotted at several dipping depth $h$ in hexadecane for the soft and stiff cantilevers. The broadening of the resonance peak shows the increase of damping as a function of $h$.}

In order to check the quality of the {measurement} and to extract a reliable value of the viscosity one has to fit the PSD in a wide frequency range. As there is no external force acting on the system, we can use the Fluctuation Dissipation Theorem (FDT)~\cite{deGroot}, {linking the mechanical} response function of the total system to the PSD {$S_{z}(f)$ of the cantilever deflection $z$}:
\begin{equation}
{S_{z}}(f)=-\frac{4k_{\mathrm{B}}T}{\omega}\mathrm{Im}(\frac{1}{G(\omega)})= \frac{4k_{\mathrm{B}}T G''}{\omega \, |G|^2}
\label{fdt.0}
\end{equation}
where $G=G'+iG''$ is the 
{the inverse of mechanical response function} of the total system, $k_{\mathrm{B}}$ is the Boltzmann constant, $T$ the temperature, $\mathrm{Im}(.)$ stands for imaginary part and $\omega=2\pi \, f$ is the angular frequency. {For a Simple Harmonic Oscillator (SHO) with viscous damping for example, the response function is~\cite{deGroot}:
\begin{equation}
G(\omega)=k-m\omega^{2}+i\gamma\omega
\label{GSHO}
\end{equation}
with $k$, $m$ and $\gamma$ respectively the stiffness, mass and damping coefficient of the SHO.}

\subsection{ A model for the damping}
The key point is to have a reliable model for $G$ and most importantly of the {damping} coefficient of this system, 
assuming that the fiber is a cylinder oscillating along its axis. As far as we know there are {no} exact analytical solutions to this problem and one has to do several approximations. The starting point is the Stokes' solution for the drag of a sphere oscillating at frequency $\omega$ in a fluid of viscosity $\eta$ and density $\rho$~\cite{landau}. Supposing that there are only geometric corrections for the fiber we can write the dissipation {$\gamma_{c}$ of the fiber-cantilever system as:
\begin{eqnarray}
\label{drag_coeff}
\gamma_{c}=\gamma_{0}(h,d)+b \sqrt{2} \frac{d}{\delta} \, \eta h= \gamma_{0}(h,d) + b d \sqrt{\rho \eta \omega} \, h
\end{eqnarray}
where $\delta=\sqrt{2\eta/(\rho\, \omega)}$ is the viscous penetration length and $b$ a generic geometric factor to be determined ($b=b(h,d)$ may depend on the values of $h$ and $d$).}
An approximated solution for the viscous coefficient {$\gamma_{0}$ {for a fully immersed rod} is~\cite{broersma}:
\begin{equation}
\label{slope}
\gamma_{0}(h,d)=\gamma_\odot + \frac{2\pi}{\ln(h/d)+\epsilon} \, \eta h
\end{equation}}
where the constant {$\gamma_\odot$} is the sum of the dissipation {in the liquid meniscus} and of the effective dissipation of the cantilever in air. The coefficient $\epsilon$ is a correction (smaller than 1), which depends on the shape of the {fiber's} cross section \footnote{ In the ellipsoid model described in ref.~\cite{broersma} and used by Xiong and \textit{al.}~\cite{Tong}, $\epsilon$ does not depend on $h$ in {our $h/d$ range} and is equal to 0.19. In a cylinder approximation,~\cite{broersma}, $\epsilon$ varies between -0.55 and -0.58 and eq.\ref{slope} is valid only for a ratio $h/d>4$. }. Except for the correction in $\ln(h/d)$, $\gamma_{0}$ has a leading linear behavior in our range of $h$.
As pointed out in ref.~\cite{landau} the term in $\sqrt{\omega}$ in eq.\ref{drag_coeff} becomes relevant only if $d>\delta$, thus it can be usually neglected at low frequencies.
Using eq.\ref{drag_coeff}, we can compute~\cite{landau} the storage {modulus $G'$ and the loss modulus $G''$ of the fiber-cantilever system:}
{
\begin{align}
G'(\omega)&=k_{c}+k_{m}-(m_{c}+m_{\mathrm{fluid}})\omega^2-b d \sqrt{\eta\rho} \ \omega^{3/2}\, h \label{real} \\
G''(\omega)&=\gamma_{0}\, \omega+b d \sqrt{\eta\rho}\ \omega^{3/2} \, h \label{imaginary}
\end{align}
where $k_{c}$ is the stiffness of the cantilever, {$k_{m}$ is the meniscus stiffness,} $m_{c}$ is the effective mass of the cantilever-fiber system, and 
$m_{\mathrm{fluid}}$ is the {mass of the displaced fluid}. In our case, $m_{\mathrm{fluid}}$ represents in the worst case of a totally immersed fiber only $3\%$ of $m_{c}$ and it will be neglected.}  {$k_{m}$ is of the order of magnitude of the liquid's surface tension, a few tens of mN/m; we will not consider its possible frequency dependance in this work.} Inserting {expressions \ref{real} and \ref{imaginary}} in eq.\ref{fdt.0} we get an expression for {$S_{z}$ that can be used to fit the measured PSD of the cantilever deflection $z$. Note} that this way we consider the cantilever-fiber system as a {SHO with a frequency dependent damping coefficient $\gamma_{c}$} {and a frequency dependent added mass that can be viewed as the mass of fluid in the boundary layer}.

{ This model, 
which takes into account the effect of the boundary layer to compute dissipation,  will be noted model  $\SHO_{B}$ 
 and the predicted spectrum $S_{z,B}^{\SHO}(f)$ will be used to fit the experimental measurement.} {Four free parameters have to be adjusted to fit the data: $k=k_{c}+k_{m}$, $m_{c}$, $\gamma_{0}$ and $\tilde{B}=bd\sqrt{\eta}$}.
 However when the resonance frequency is small enough (i.e. $d<\delta$ ) then we may {impose {$\tilde{B}=0$}. Thus, in this case, we recover the classic $\SHO$ model, noted  $\SHO_0$ (i.e. $\tilde B=0$), leading to the spectrum $S_{z,0}^{\SHO}(f)$ 

\subsection{Simple Harmonic Oscillator}
We consider first the SHO model for the {soft cantilever. Indeed at the resonance frequency ($\SI{8.2}{kHz}$), $\delta$ varies between $\SI{6.8}{\mu m}$ for dodecane and $\SI{28}{\mu m}$ for silicon oil v20, thus for all the liquids that we use $\delta>d\simeq \SI{3}{\mu m}$ and $\tilde{B}$ can be neglected. 
We therefore fit the spectra using $S_{z,0}^{\SHO}(f)$ which has the advantage of having only {three free parameters}.}

In order to measure $\gamma_{0}$ we proceed in the following way. To begin with, we simply fit the resonant frequency peak, as done in ref.~\cite{Tong}, for each dipping depth $h$ and each liquid. This method is not very precise because 
the values of $\gamma_{0}$ depend on the chosen {fitting range} around the {soft} resonance. {Therefore, one has to decide a criterium to select this fitting range, which, in our case,  is chosen  in order to have the best fit in the largest part of the spectra} 
The results of these fits around the resonance peak are shown in fig.\ref{profondeurs}. We notice that the spectra {$S_{z,0}^{\SHO}$} do not fit properly our data at low frequency where {some extra noise is present below $\SI{2}{kHz}$}. The use of {one} more free parameter {with model $\SHO_B$} does not improve the {quality of the fitting, because the additional terms deform the $\SHO_B$ noise spectrum even further away from our data}. 

{In order to understand where the noise increase at $f<\SI{2}{kHz}$ is coming from,} we analyze more precisely the spectra in fig.\ref{profondeurs}. We notice that the spectrum in air presents another resonance at about $\SI{29}{kHz}$, which corresponds to the first flexion mode of the fiber oscillations. This resonance disappears when the fiber is immersed. To {find out what happens} to this resonance, we decided to do the same measurement with {the same kind of fiber (length and diameter) on a more rigid cantilever having a resonance around $\SI{60}{kHz}$}. The aim is to {avoid damping the fiber noise in the inertial tail of the resonance}. The spectra measured in air and in hexadecane using this stiffer cantilever-fiber system are plotted in figure \ref{HFhexadecane}. In the spectrum in air, we clearly see  {the resonance of the cantilever at \SI{57}{kHz}} and the resonance of the fiber at \SI{24}{kHz}, at} approximately the same frequency than the {previous system}. When the fiber is dipped into the fluid this resonance becomes over-damped and the cut-off frequency goes towards low frequencies close to $\SI{2}{kHz}$. {In contrast}, the resonance of the {stiff} cantilever behaves similarly to that of the {soft one}, presenting a continuous broadening of the resonance peak as a function of $h$. All these remarks suggest that there is probably a coupling between the fiber and the cantilever oscillations which {perturbs} the simple picture of {a} SHO.

\subsection{Coupled oscillators}
Let us develop a model of Coupled Harmonic Oscillators (CHO) {for the cantilever (deflection $z$) and the fiber (deflection $y$ of its extremity).
{In a first approximation, the} cantilever motion corresponds to a forcing of the clamping base of the fiber along its axis, thus we will neglect the effect of the cantilever deflection on that of the fiber. The fiber is thus modeled as a classic SHO and in the Fourier transform $y_\omega$ of its deflection $y$ is described by:
\begin{equation}
\label{motion.fiber}
(k_{f}-m_{f}\omega^2+i \gamma_{f} \omega) y_\omega = F_{f}
\end{equation}
where $m_{f}$, $\gamma_{f}$ and $k_{f}$ are respectively the effective mass, dissipation and stiffness of the fiber and $F_{f}$ is a} delta correlated thermal noise acting on the fiber, whose spectrum is $S_{F_{f}}=4k_BT \gamma_{f}$.

{The fiber motion translates into a torque of the cantilever end, thus the equation describing the Fourier transform $z_\omega$ of  $z$ is coupled to the fiber deformation $y$:
\begin{equation}
\label{motion.cantilever}
(G'+iG'')z_\omega=F_{c}+\alpha y_\omega
\end{equation}
where $G'$ and $G''$ are defined in eqs.~\ref{real} and \ref{imaginary}, and $F_{c}$ is a delta correlated thermal noise, whose spectrum is $S_{F_{c}}=4k_BT \gamma_{c}$. The term $\alpha y$ (with $\alpha$ {the} coupling coefficient) assumes the simplest coupling with the deflection of the fiber $y$. The PSD $S_{z}^{\CHO}(f)$ of $z$} can be computed from eqs. \ref{motion.fiber} and \ref{motion.cantilever} by making the very reasonable hypothesis that {$F_{f}$ and $F_{c}$} are uncorrelated noise. We get:
{\begin{equation}
\label{coupled.spectra}
S_{z,n}^{\CHO}(f)=\frac{4k_{\mathrm{B}}T}{|G|^2}\left(\gamma_{c}+\frac{\alpha^2\gamma_{f}}{(k_{f}-m_{f}\omega^2)^2+(\gamma_{f}\omega)^2}\right)
\end{equation}
{where  $n$ stands for either $0$ or $B$ depending  whether we impose $\tilde{B}=0$ or not. Eq.\ref{coupled.spectra} can be written in a more compact form:
\begin{equation}
\label{coupled.spectra.simple}
S_{z,n}^{\CHO}(f)=S_{z,n}^{\SHO}(f)\left(1+\frac{\alpha^2}{\gamma_{c}} \frac{S_{y}(f)}{4k_{\mathrm{B}}T}\right)
\end{equation}
where $S_{y}$ is the PSD of the fiber thermal noise.} When the fiber is dipped into the fluid, we notice that the motion of the fiber is over-damped. We can thus simplify and set: 
{\begin{equation}
\label{fiber.spectra.simple}
S_{y}(f)=\frac{4k_{\mathrm{B}}T\tau_f}{k_{f}(1+(\tau_f\omega)^2)}
\end{equation}
where $\tau_f=\gamma_{f}/k_{f}$} is the relaxation time of the fiber. 

We can try to fit our data with these $\CHO_n$ models}, i.e. eq.
\ref{coupled.spectra.simple} and \ref{fiber.spectra.simple}. Because of the large number of parameters in the model we proceed in the following way using first the model $\CHO_0$ with $\tilde{B}=0$. 
We begin to fit the spectrum around the cantilever resonance to estimate $\gamma_{0}$ and to obtain a first approximation of {$S_{z,0}^{\SHO}$}. Inserting this first approximation in eq.~\ref{coupled.spectra.simple}, we can fit the expression {$S_{z}^{\meas}/S_{z,0}^{\SHO}-1$} with a Lorenztian, from which the values of {$k_{f}/\alpha^2 $} and $\tau_f$ are obtained. Using these values in 
eq.\ref{coupled.spectra.simple} one can improve the fit of {$S_{z,0}^{\SHO}$} and repeat the iteration. After 3 iterations the values of the parameters become stable and we obtain a good fit on the whole frequency range (figure \ref{HFhexadecane}). The {$S_{z,0}^{\CHO}$} fits well the resonant peak as can be seen in the inset of fig.\ref{HFhexadecane}. However we see that the fit is not correct around $\SI{10}{kHz}$ where the fitting curve is systematically above the data. The problem could come from the fact that we perform the fit around the resonance at $\SI{60}{kHz}$  keeping $\tilde B=0$. At such a high frequency the {$\CHO_0$ model} is probably not adequate because the boundary layer thickness $\delta$ is about \SI{4.5}{\mu m}, which is close to the fiber diameter. Thus one should use the {$\CHO_B$ model} which takes into account the boundary layer effects.

\begin{figure}[htb]
\begin{center}
\includegraphics[width=0.8\linewidth]{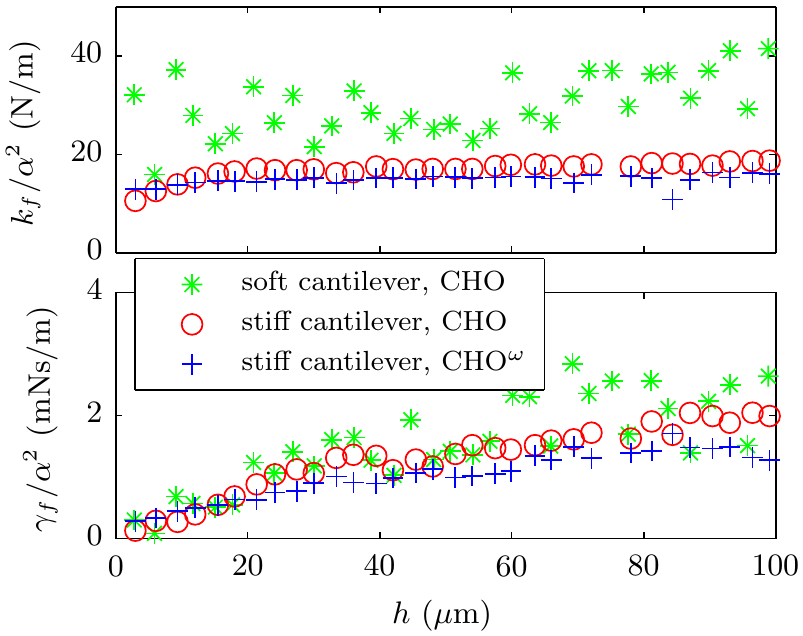} 
\caption{{Evolution of the fiber stiffness $k_{f}$ and damping coefficient $\gamma_{f}$ as a function of the dipping depth $h$ of the fiber in hexadecane. The stiffness is constant, while the damping increases from 0 and saturates for large $h$, where immersion approaches the base of the fiber.}}
\label{stiffHF}
\end{center}
\end{figure}


{To fit the data with {$S_{z,B}^{\CHO}(f)$}, we use the same iteration approach with one difference:
to reduce the number of free parameters, the stiffness and mass are now fixed to the values measured in air, $k_{c}^{\mathrm{air}}= \SI{3.7}{N/m}$ and $m_{c}^{\mathrm{air}}= \SI{2.9e-11}{kg}$. This is justified because the other contributions to the stiffness and mass are negligible with respect to the values in air in the high frequency experiment.} After several iterations we get a correct fit of the data except around the resonance (see inset figure \ref{HFhexadecane}) where the fit {$S_{z,B}^{\CHO}(f)$} is higher than the data. This indicates that this model {for the cantilever's resonance} at high frequency is not perfect. Thus we conclude that in order to discriminate between the two models {($\tilde{B}=0$ and $\tilde{B}\ne0$)}, we would need {an even larger} frequency range. One could check at low frequencies, where {$S_{z,B}^{\SHO}(f)$}  {has} a dependence in $\sqrt{\omega}$ instead of the flat curve of {$S_{z,0}^{\SHO}(f)$}. But in our data, the low frequency part is hidden by {$S_{y}(f)$}. On the other hand one could use the high frequency part of the spectra. Indeed at high frequency, {$S_{z,B}^{\CHO}(f)$} should decrease in $\omega^{-7/2}$ instead of $\omega^{-4}$ of {$S_{z,0}^{\SHO}(f)$}. But in this case the intrinsic noise of the interferometer {hides} the data and this comparison is not possible. {Besides, we could see in air the second mode of the fiber very close to the cantilever's resonance on most high frequency probes we tested, so that this mode may also disturbs the cantilever's resonance. As a result, it is not possible to clearly separate the two contributions to the dissipation {$\gamma_{c}$} from our measured noise spectra.}

We can see in fig.\ref{stiffHF} the evolution of {the} stiffness {$k_{f}$} and {the} dissipation {$\gamma_{f}$} of the fiber lateral oscillations as a function of dipping depth estimated from the two models. It is interesting to notice that the values obtained from the two models are very close and have a dependence on $h$ which is quite reasonable. Except at the very beginning \footnote{when the fiber is just touching the liquid, {surface tension effects may perturb the {measurement}}}, the stiffness is constant as expected. Instead, the dissipation increases and tends to saturate at large $h$. This behavior can be understood considering that {$\gamma_{f}$} is the damping of the {deflection} mode of the fiber which has a large displacement at the free extremity and a small one towards the anchoring point. Therefore the non moving part of the fiber will not contribute to dissipation and {$\gamma_{f}$} increases fast at small $h$ (region of large lateral displacement) and saturates above a certain $h$ (region of small {displacement}). 
{ From this data it is possible to have the order of magnitude  of  $\alpha$. Indeed taking into account the size  of the fiber and its Young modulus one estimates $k_f \simeq 0.1 N/m $ and $\alpha \simeq 0.06 - 0.07$ which is a reasonable value for the coupling} 



\begin{figure}[htb]
\begin{center}
\includegraphics[width=0.8\linewidth]{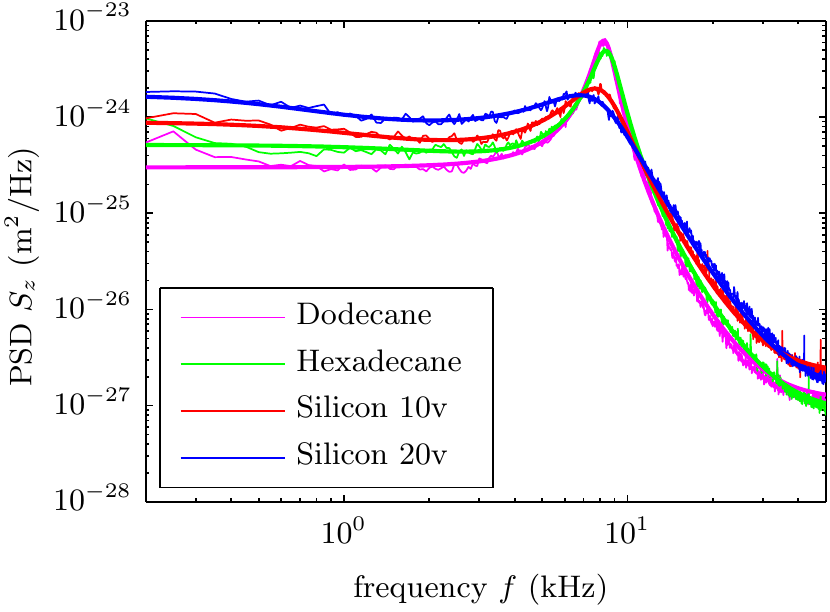}
\caption{{Soft cantilever, dipping depth about \SI{63}{\micro m}: (thin lines) PSD $S_{z}^{\meas}(f)$ of the measured thermal noise for each liquid ; (thick lines) fit with $S_{z,0}^{\CHO}(f)$, modeling the coupled oscillators.}\label{pviscosite}}
\end{center}
\end{figure}

{Looking at fig.\ref{profondeurs} we notice that the resonance of the fiber in air is about at the same frequency for the soft probe as for the stiffer one}  \footnote{{In fig.\ref{profondeurs}} the amplitude {of} the fiber resonance is very low because being at a frequency larger than the cantilever resonance, it is strongly filtered by the cantilever response (see eq.\ref{coupled.spectra})}. We can thus suppose that the cut-off frequency when the fiber is immersed is also around \SI{2}{kHz}, that is to say, at the left of the cantilever resonance. Therefore we can use the same iteration than before to analyze our spectra using eq.\ref{fiber.spectra.simple} for the spectrum of the fiber and {model $\CHO_0$}. On figure \ref{pviscosite}, we can see fits for each liquid at $h=\SI{63}{\mu m}$. We see that now the model fits our data properly on the whole frequency range. 
We can also see on fig.\ref{stiffHF} the same evolution for the stiffness and the dissipation of the fiber. Data are noisier than those at high frequency because the resonance of the cantilever is {closer} to the cut-off frequency of the fiber. 

\subsection{Viscosity measurement at low frequency}
{In order to perform viscosity measurements, we choose to focus on low frequency measurements, which present several benefits when compared to high frequency measurements. Firstly the hanging-fiber probes are less difficult to fabricate on softer and longer cantilevers. Secondly, the signal over noise ratio is much higher with soft cantilevers. Finally, the viscosity measurement is based on the dissipation $\gamma_{0}$, and requires neither the calibration of the unknown coefficient $b$ not the knowledge of the liquid's density $\rho$.}

\begin{figure}[htb]
\begin{center}
\includegraphics[width=0.8\linewidth]{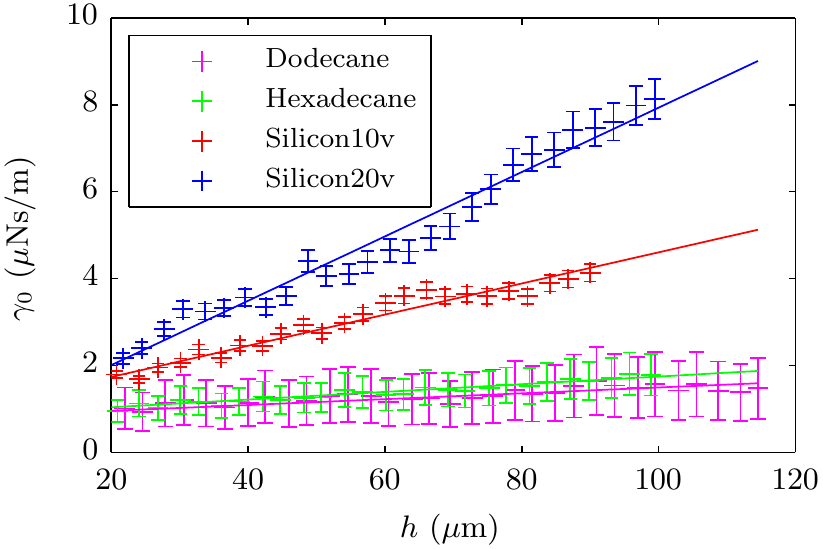} 
\includegraphics[width=0.8\linewidth]{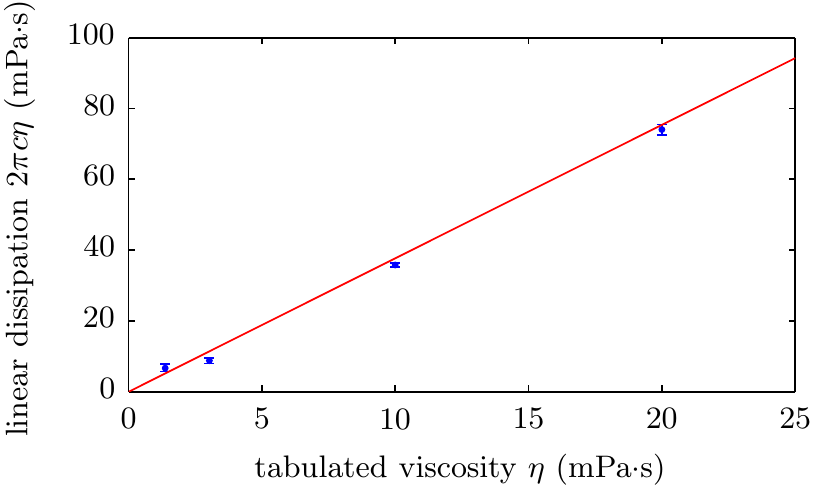}
\end{center}
\caption{a) Dissipation {$\gamma_{0}$} as a function of $h$ for each liquid. b)
Linear dissipation of each liquid ($2\pi c\eta$) as a function of the tabulated viscosity for alcanes and silicon oils. 
}
\label{h_liquid}
\end{figure}

{We extract the dissipation coefficient $\gamma_{0}$ for each liquid at each depth from the $\CHO_0$ fitting procedure on the soft probe's data}. The results are plotted in fig.\ref{h_liquid}a), where we can see a linear behavior as a function of $h$ {in the displayed immersion range where $h/d > 6$}. {The error bars take into account the standard deviation of the mass and the stiffness of the cantilever during the dipping.} Our range in $h$ is too small to measure the logarithmic correction of eq.\ref{slope} and we rewrite it as: {$\gamma_{0}=\gamma_\odot + 2\pi c \eta h${, where $c$ is a parameter that depends on the geometry of the probe}.} 
From the plot of fig.\ref{h_liquid}a), we extract the slope $2\pi c \eta$ for each liquid, and plot it in fig.\ref{h_liquid}b) as a function of the tabulated viscosity $\eta$. The data are aligned on a straight line which correctly crosses the (0,0) point with a slope {$2 \pi c=3.67 \pm 0.07$. We thus obtain $c=0.58 \pm 0.01$}. 

This value is interesting because it clearly excludes that $\gamma_{0}$ is simply given by eq.\ref{slope}. Indeed for our measuring range {$6<h/d<30$} and for $-1<\epsilon<0$ one finds that the function {$h/(\ln(h/d)+\epsilon)$} is well approximated by a straight line $(c' \, h+c_o' )$ with $c'$ always very close to $0.2$, which is incompatible to the measured value of $c$. {This difference is not so surprising since the model in ref.~\cite{broersma} considers a fully immersed fiber whereas it crosses the liquid-air interface in our system, and a small transverse component in the driving by the cantilever also contributes to the measured dissipation.} Thus 
it is necessary to calibrate $c$ for each fiber in order to have reliable results.

\section{Conclusion}
In this article, we have presented a passive measure of the local viscosity at micrometer scale for a range of viscosity between \SI{1}{mPa.s} and at least \SI{20}{mPa.s}, with 5\% of accuracy for its absolute value. 
We have proposed a model which takes into account the fiber-cantilever coupling, and have shown that this model fits the {thermal noise} spectra. 
By increasing the resonant frequency of the cantilever, we {could} clearly separate the frequency of the first mode of the cantilever and the first mode of the fiber. This increases the quality of the data {for the fiber's mode but does not simplify the viscosity measurements from the cantilever's resonance} 
{We have tested a model taking into account the frequency dependent boundary layer terms (model $\SHO_B$), but couldn't discriminate it from a simpler $\SHO_0$ model by fitting the measured spectra, despite our high resolution deflection sensor. As a result, we prefer using the simpler model with less free parameters. Finally, our data {gives insights into the dissipation mechanism of a micro-fiber partially immersed in a viscous fluid}.}
 
{This microrheology} method based on {a} {dipping} fiber has several advantages. It does not require a transparent liquid, and only a very small amount of liquid is needed.
{The accessible viscosity range is limited at \SI{1}{mPa.s} in our measurements because for lower viscosities the fiber's mode becomes to close to the cantilever's mode and cannot be separated from it. The upper viscosity bound has not been reached yet, since over-damped motion for higher viscosities can still be analysed}.

{This method calls nevertheless for} some precautions. Using functionalized cantilevers is often a good idea but it is necessary to check {the} influence of added elements~\cite{justine}. Here, the fiber-cantilever coupling can strongly affect the accuracy of the results. {Finally, we assumed a no-slip boundary condition on the fiber, which is justified because the intrinsic slip length remains below \SI{30}{nm} for simple liquids \cite{audrey} (while it reaches micrometric scale for polymer melts \cite{leger}). Beware that boundary slippage can strongly affect the $b$ and $c$ coefficients when working with nano-fibers like in ref.~\cite{ondarcuhu,yazdanpanah} or when studying polymer melts with micro-fibers as in \cite{sauer}}.

\section{Acknowledgments}
This paper has been supported by the ERC project OUTEFLUCOP.


\begin{thebibliography}{}

\bibitem{Mason} T. M Squires and T. G. Mason, Annu. Rev. Fluid Mech. 42, 413 (2010).

\bibitem{Gittes} F.Gittes, B. Schnurr, P.D. Olmsted, F.C. MacKintosh, C.F. Schmidt, Phys. Rev. Lett. 79, 3286 (1997). 

\bibitem{Solano} J. R. Gomez-Solano, A. Petrosyan and S. Ciliberto
Europhysics Lett., 98 (2012) 10007

\bibitem{SMR_viscosity}
I. Lee, K. Park, and J. Lee, 
{\it Rev. Scien. Inst.}, {\bf 83}, 116106 (2012). 

\bibitem{Tong}
X. Xiong, S. Guo, Z. Xu, P. Sheng, P. Tong,
{\it Physical Review E}, {\bf 80}, 061604 (2009).

\bibitem{Paolino-2013}
P.~Paolino, F.~Aguilar Sandoval, and L.~Bellon, 
{\it Rev. Sci. Instrum.}, {\bf 84}, 095001 (2013).

\bibitem{Schonenberg-RSI-89}
C.~Schonenberg, and S.~F.~Alvarado, 
{\it Rev. Scien. Inst.}, {\bf 60}, 3131 (1989).

\bibitem{2002Bellon}
L.~Bellon, S.~Ciliberto, H.~Boubaker, and L.~Guyon, 
{\it Opt. Comm.}, {\bf 207}, 49 (2002).

\bibitem{deGroot}
S.R. de Groot and P. Mazur, {\it Non equilibrium thermodynamics}, Dover(1984).

\bibitem{broersma}
S. Brorsma, 
{\it J. Chem. Phys.}, {\bf 32}, 1632 (1960). 

\bibitem{landau} L.
D. Landau and E.M. Lifshitz, Fluid Mechanics (Pergamon Press, London,1966),

\bibitem{justine} J. Laurent, A. Steinberger, L. Bellon, 
{\it Nanotechnology}, {\bf 24}, 225504 (2013)

\bibitem{audrey} Bouzigues C.I., Bocquet  L., Charlaix E., Cottin-Bizonne  C., Cross B., Joly L., Steinberger A., Ybert C. and Tabeling P.,
{\it Phil. Trans. R. Soc. A}, {\bf 366} (2008) 1455.

\bibitem{leger} L\'eger L., Hervet H., Massey G. and Durliat E.,
{\it J. Phys.: Condens. Matter}, {\bf 9 }(1997) 7719. 

\bibitem{ondarcuhu} Delmas M., Monthioux M. and  Ondar?uhu T.,
{\it Phys. Rev. Lett.}, {\bf 106 }(2011) 136102. 

\bibitem{yazdanpanah} Yazdanpanah M .M., Hosseini M., Pabba S., Berry S. M., Dobrokhotov V.V., Safir  A., Keynton R. S. and Cohn R.W.,
{\it langmuir}, {\bf 24 }(2008) 13753. 

\bibitem{sauer} Sauer B. B. and Kampert W. G.,
{\it Colloid Interface}, {\bf 199 }(1998) 28. 



\end{thebibliography}
\end{document}